\documentclass[%prl,
aps,showpacs%preprint %tightenlines
 ,twocolumn
]{revtex4}
\usepackage{tikz}
\usepackage{adjustbox}
\usetikzlibrary{arrows, hobby}
\usepackage{wrapfig}
\usepackage{graphicx}
\usepackage{amsmath}
\usepackage{amssymb}
\usepackage{hyperref}
\usepackage{ mathrsfs }
\usepackage{indentfirst}
\usepackage{braket}
\usepackage{color}
\usepackage{empheq}
\usepackage{lipsum,adjustbox}
\usepackage[outdir=./]{epstopdf}
\usetikzlibrary{calc,positioning}
\tikzstyle{beamsplitter}=[fill=blue, fill opacity=0.2]

\begin{document}

\title{Quantum Metrology in the Kerr Metric}

\author{ S. P. Kish and T. C. Ralph}\affiliation{Centre for Quantum Computation and Communication Technology, \\
	School of Mathematics and Physics, University
	of Queensland, Brisbane, Queensland 4072, Australia}

\date{\today}

\begin{abstract}
	{}
	A surprising feature of the Kerr metric is the anisotropy of the speed of light. The angular momentum of a rotating massive object causes co- and counter-propagating light paths to move at faster and slower velocities, respectively as determined by a far-away clock. Based on this effect we derive ultimate quantum limits for the measurement of the Kerr rotation parameter $a$ using an interferometric set up. As a possible implementation, we propose a Mach-Zehnder interferometer to measure the ``one-way height differential" time effect. We isolate the effect by calibrating to a dark port and rotating the interferometer such that only the direction dependent Kerr-metric induced phase term remains. We transform to the Zero Angular Momentum Observer (ZAMO) flat metric where the observer see $c=1$. We use this metric and the Lorentz transformations to calculate the same Kerr phase shift. We then consider non-stationary observers moving with the planet's rotation, and find a method for cancelling the additional phase from the classical relative motion, thus leaving only the curvature induced phase. 
\end{abstract}


\maketitle

%\tableofcontents

\vspace{10 mm}
\section{Introduction}
%{\color{red} How does the Michelson interferometer look like in rotating Earth frame? Include diagram? Change subheadings. Microwave resonator experiment $\Delta c/c \approx 10^{-17}$. Relate to other experiments? Is it possible with current day technologies?}
The noise induced by a measurement device is fundamentally restricted by limits set by quantum mechanics. Quantum metrology is the study of these lower limits for the estimation of physical parameters \cite{GIO11}. 
Techniques in quantum metrology can assist in developing devices to measure the fundamental interplay between quantum mechanics and general relativity at state-of-the-art precision. A prime example is the detection of gravitational waves from black-hole mergers by LIGO \cite{ABB16}.

Recently there have been investigations of how we can exploit quantum resources to measure space-time parameters such as the Schwarzschild radius $r_s$ and the Kerr parameter $a$ in the rotating Kerr metric \cite{BRU14, MILB, qbits, BRU13}. Quantum communications were shown to be affected by the rotation of Earth \cite{kohlrus}. However, more fundamental effects in general relativity induced by the Kerr metric were not analysed. One interesting feature of the Kerr metric is the anisotropy of the velocity of light (null geodesics). The rotating massive object causes co- and counter- propagating light to move at faster and slower velocities, respectively. 

In this paper, we note that there is a phase shift of co-moving light beams at different radial positions in the Kerr metric. We use a Mach-Zehnder (MZ) interferometer to probe for this phase. We isolate the effect by calibrating to a dark port and rotating the interferometer and due to the anisotropy of $c$, only the Kerr phase term remains. From this, we can construct using Quantum Information techniques lower bounds for the variance of parameter estimation of $a$ \cite{BRU14, BRU13, cram}. 

Locally, we can find a co-rotating frame in which the space-time is locally flat (``the zero angular momentum ring-riders") \cite{visser}. We find that the locally measured velocity of light is $c=1$ as expected in the flat metric. If an observer Alice compares the locally measured time with Bob who is a ring-rider at a different radius, there will be a disagreement of simultaneity of events. We also consider non-stationary observers that are moving in the rotational plane of Earth. As expected, we find an additional phase term from rotation and special relativistic time dilation. We find that this term is dominant compared to the Kerr phase. Finally, we compare the magnitude of the Kerr phase on Earth to that acheivable by microwave resonator experiments \cite{herm}.

This paper is organized as follows. We first introduce the full Kerr metric in Section \ref{kerr1} for a rotating black hole. In Section \ref{approximatekerr}, we approximate the Kerr metric to first order in angular momentum where the mass quadrupole moment for massive planets or stars is dropped in the weak field limit. In Section \ref{far}, we solve for the null geodesic to determine the velocity of light in the equatorial plane. We find the anisotropy in $c$. Next in Section \ref{height}, we calculate the ``height differential effect" which could be detected by a Mach-Zehnder interferometer above a massive planet.  

In Section \ref{qlimit}, we determine quantum limits of the Kerr space-time parameter $a$ for the height differential effect. In Section \ref{mz}, we focus on the stationary Mach-Zehnder interferometer in the weak field limit and calculate the phase shift. We comment on how we can calibrate to a dark port and rotate the interferometer to isolate the Kerr phase. We compare the magnitude of the Kerr phase with the Schwarzschild phase for Earth parameters. In Section \ref{zamo}, we use the co-moving flat metric in which the so-called ``ring-rider" measures $c=1$. In Section \ref{ringriderper}, we demonstrate an alternative calculation using Lorentz transformations between stationary and ring-riders to find the phase detected at the output of the MZ interferometer. We also confirm that the ``two-way" velocity of light is $c=1$ as detected by a Michelson interferometer at rest in the Kerr metric. Furthermore, we consider the motion of non-stationary observers on the rotating planet. In Section \ref{extremal}, we consider an extremal black hole and we numerically find the full strong field solution of the Kerr phase. Finally, we conclude by commenting on the feasibility of detecting the light anisotropy.

\section{Kerr Rotational Metric}
\label{kerr1}
The metric describing the space-time of an axially symmetric rotating massive body is given by the Hartle-Thorne metric, which includes the dimensionless mass quadrupole moment $q$ and the angular momentum $j$ of the massive body \cite{hartle, alfaro}. The mass quadrupole moment $q=k j^2$ where $k$ is a numerical constant that depends on the structure of the massive body. The Kerr metric for a black hole is obtained from the Hartle-Thorne metric by setting $q=-j^2$ and transforming to Boyer-Lindquist coordinates \cite{abram, bini}.

A rotating black hole tends to drag the space-time with its rotation. The Kerr metric used to describe this space-time includes the Kerr rotation parameter ``$a$" which quantifies the amount of space-time drag. The Kerr line element in Boyer-Lindquist coordinates $(t,r,\theta,\phi)$ is \cite{visser,hartle, taylor}:

\begin{equation}
\begin{split}
ds^2&=-(1-\frac{r_s r}{\Sigma})dt^2+\frac{\Sigma}{\Delta} dr^2+\Sigma d\theta^2 \\
&+(r^2+a^2+\frac{r_s r a^2}{\Sigma} \sin^2{\theta})\sin^2{\theta} d\phi^2-\frac{2 r_s r a \sin^2{\theta}}{\Sigma} d\phi dt
\end{split}
\label{kerrmetric}
\end{equation}

Where $\Delta:=r^2-r_s r+a^2$, $\Sigma:=r^2+a^2 \cos^2{\theta}$ and $a=\frac{J}{M c}$ where $J$ is the angular momentum of the black hole of mass $M$. Note that the Schwarzschild radius $r_s=\frac{2GM}{c^2}\equiv 2 M$ where we work in natural units for which $c=1$ and $G=1$. Compared with the Schwarzschild metric, the cross term $dt d \phi$ introduces a coupling between the motion of the black hole and time, which leads to interesting effects. 

When $r_s=0$, the space-time is flat and reduces to $ds^2=- dt^2+\frac{1}{1+\frac{a^2}{r^2}} dr^2+(r^2+a^2) d\phi^2$. At first glance, this metric doesn't seem flat. However, we have used the oblong sphere coordinates $x=\sqrt{r^2+a^2} \sin {\theta} \cos{\phi}$, $y=\sqrt{r^2+a^2} \sin {\theta} \sin{\phi}$ and $z=r \cos{\theta}$. 

\subsection{Approximate Kerr metric for rotating massive bodies}
%The full exterior metric of a rotating axially symmetric massive body is given by the Hartle-Thorne metric which includes the dimensionless mass quadrupole moment $q$ that depends on the structure of the body \cite{hartle, alfaro}. The more commonly discussed Kerr metric describes the exterior metric outside of a black hole with angular momentum $a=\frac{J}{M}$ to second order \cite{abram, bini}. 
\label{approximatekerr}
The mass quadrupole moment of a {\it massive planet} is proportional to the angular momentum squared. Thus, we cannot use the Kerr metric in Eq. \ref{kerrmetric} where the proportionality constant for {\it black holes} is $k=-1$. 
However, in the weak field limit $a<<r$, we can truncate the Kerr metric to first order in $\frac{a}{r}$. Thus the approximate Kerr metric is given by:
\begin{equation}
\begin{split}
ds^2&=-(1-\frac{r_s}{r}) dt^2+ (1-\frac{r_s}{r})^{-1} dr^2+r^2 d\theta^2 \\
&+r^2 \sin^2{\theta} d\phi^2-\frac{2 r_s a \sin^2{\theta}}{r} d\phi dt
\end{split}
\label{approximate}
\end{equation}

This approximate Kerr metric disregards the mass quadrupole moment of the massive body. It is equivalent to the Hartle-Thorne metric with the same first order approximation \cite{alfaro2}. When we refer to a massive planet or star, we will use this approximate Kerr metric. We wish next to determine the tangential velocity of light close to the massive object as seen by a far-away observer.

%Let's compare this approximate Kerr metric with the Hartle-Thorne metric. The approximate Hartle-Thorne metric is given by \cite{alfaro2}:
%
%\begin{equation}
%\begin{split}
%ds^2&=-(1-\frac{r_s}{r}+\frac{r_s q}{r^3} P_2 (\cos{\theta}))dt^2\\
%&+(1+\frac{r_s}{r}+ \frac{r_s^2}{r^2} - 2 \frac{2 r_s q}{r^3} P_2 (\cos{\theta})) dr^2 \\
%& + r^2(1-\frac{2 q r_s}{r^3} P_2(\cos{\theta})) (d\theta^2+\sin^2{\theta} d\phi^2) \\
%&-\frac{2 r_s j}{r} \sin^2{\theta} dt d\phi
%\end{split}
%\end{equation}
%Where $P_2(\cos{\theta})=(3 \cos^2{\theta}-1)/2$. Thus if we set $q=k j^2=\frac{k}{4} r_s^2 a^2$, and truncate to first order in $\frac{a}{r}$. 

\subsection{Far-away velocity of light}
\label{far}
In the equatorial plane (where $\theta=\frac{\pi}{2}$), for the null light geodesic, we set $ds^2=0$ and determine the solution for the tangential velocity of light according to Kerr time coordinate $t$. The Kerr time coordinate corresponds to a clock from the gravitating massive body hence this is the speed of light inferred by a far-away observer. Using Eq. \ref{approximate}:
 
\begin{equation}
\begin{split}
ds^2&=0=-(1-\frac{r_s}{r}) dt^2+ (1-\frac{r_s}{r})^{-1} dr^2 \\
&+r^2 d\phi^2-\frac{2 r_s a}{r} d\phi dt
\end{split}
\end{equation}

   %The definition of the measured radius differs from the Schwarzschild metric. The reduced circumference $2 \pi R$ where $R=\sqrt{r^2+a^2+\frac{r_s}{r} a^2}$ defines the measured radius by an observer from which $r$ can be extracted. 
   The tangential distance is $d x=r d \phi$ and the light geodesic solution is: 
\begin{equation}   
   0=-(1-r_s/r) +\dot{x}^2-\frac{2 r_s a}{r^2} \dot{x}
   \end{equation}
   Where $\dot{x}=\frac{d x}{dt}$. However, if $\frac{a}{r}<<1$ and $\frac{r_s}{r}<<1$ we have the weak field solution:
%   \begin{equation}
%   \begin{split}
%    \frac{d x}{dt}=\frac{r_s a}{r^2} \pm \sqrt{\frac{ r_s^2 a^2}{r^4}+(1-\frac{r_s}{r})}
%    \end{split}
%    \label{full}
%   \end{equation}

     \begin{equation}
     \begin{split}
     \frac{d x}{dt} &\approx \frac{r_s a}{r^2} \pm \sqrt{1-\frac{r_s}{r}} \\
     &\approx \pm (1-\frac{r_s}{2 r} \pm \frac{r_s a}{r^2})
     \end{split}
     \label{dxdt}
     \end{equation} 
   
Where we have used the Taylor expansion $\sqrt{1-x}\approx 1-\frac{x}{2}$. We have two solutions representing counter- and co-rotating light. Notice that locally, $\frac{dx}{dt} \frac{dt}{d\tau}\approx (1-\frac{r_s}{2 r}+\frac{r_s a}{r^2})(1+\frac{r_s}{2r}) =1+\frac{r_s a}{r^2}$ can exceed $1$ for the positive solution. However, we cannot naively use the Schwarzschild co-ordinate time in this curved metric. Later we will show that there is a locally flat metric where $c=1$.

\subsection{Height differential effect}
\label{height}
Let's consider a stationary observer in the Kerr metric sending co-moving beams of light that travel tangentially at velocities $c_1=1-v_1-\frac{r_s}{2 r_1}$ and $c_2=1-v_2-\frac{r_s}{2 r_2}$ at radiuses $r_1$ and $r_2=r_1+h$ where $h$ is the coordinate height. For simplicity we made the weak field approximation and only retained terms from Eq. \ref{dxdt} to first order in $v_{1,2}=\frac{r_s a}{r^2_{1,2}}$. The light travels the distance $L$ with time $t_1=\frac{L}{c_1}$. Similarly, the second observer measures the travel time $t_2=\frac{L}{c_2}$. The far-away observer agrees that the length $L$ is the same for both. Thus the time delay to first order is 
\begin{equation} 
\begin{split}
\Delta t_r&=\frac{L}{c_1}-\frac{L}{c_2}=L (\frac{1}{(1-v_1-\frac{r_s}{2r_1})}-\frac{1}{(1-v_2-\frac{r_s}{2r_2})}) \\
&\approx L (r_s a (\frac{1}{r_1^2}-\frac{1}{r^2_2}))+\frac{L h r_s}{2 r_1 r_2}\\
&\approx \frac{L r_s a h (2 r_1 +h)}{r_1^4 (1+\frac{h}{r_1})^2} +\frac{L h r_s}{2 r_1^2 (1+\frac{h}{r_1})} \\
&\approx  \frac{2 L r_s a h}{r_1^3} +\frac{L h r_s}{2 r_1^2}
\end{split}
\label{height1}
\end{equation}
Where we have ignored the cross term $\frac{r_s v_1}{2r_1}-\frac{r_s v_2}{2 r_2}$ since it is much smaller and enforced the approximation $h << r_1$. This time delay can be incorporated into a Mach-Zehnder interferometric arrangement which can be rotated along its centre to measure the phase for $+a$ and $-a$ as will be discussed shortly.

\section{Quantum Limited Estimation of the Kerr space-time parameter}
\label{qlimit}
Using these time delays, we want to determine the ultimate bound for estimating the Kerr metric parameter $a$. The variance of an unbiased estimator is determined by the Quantum Cramer-Rao (QCR) bound \cite{cram}. In quantum information theory, for $M$ number of independent measurements, the QCR bound for the linear phase estimator $\phi$ is given by $\braket{\Delta \hat{\phi} ^2} \ge \frac{1}{M H (\Delta \phi)}$. Where $H(\phi)$ is the Quantum Fisher Information which quantifies the amount of information obtainable about a parameter using the optimal measurement. 
%Therefore, for the parameter $a$ we measure the phase $\Delta \phi_{Total}=\omega \Delta t_{Total}$ using a Michelson interferometer to determine the two-way time delay of light. The QCR bound for the standard deviation of $a$ is:

%\begin{equation}
%\frac{\braket{\Delta a}}{a}\ge \frac{1}{a \frac{d (\Delta \phi)}{d a}\sqrt{M H(\phi)}}\approx\frac{r^2 R^2}{\omega L a^2 r_s^2 \sqrt{M H(\Delta \phi)}}
%\end{equation}

%And for the Schwarzschild radius, we similarly have:
%\begin{equation}
%\frac{\braket{\Delta r_s}}{r_s}\ge \frac{1}{r_s \frac{d (\Delta \phi)}{d r_s}\sqrt{M H(\phi)}}\approx\frac{r^2 R^2}{ \omega L a^2 r_s^2\sqrt{M H(\Delta \phi)}}
%\end{equation}

We have seen that we can measure the phase $\Delta \phi=\omega \Delta t_r$ at different heights where $\omega$ is the central frequency of the probe and $\Delta t_r$ is given by Eq. \ref{height1}. The QCR bound for the Kerr rotation parameter is then:

\begin{equation}
\frac{\braket{\Delta a}}{a}\ge \frac{r_1^3 (1+\frac{h}{r_1})^2}{\omega L a r_s h (2+\frac{h}{r_1})\sqrt{M H(\Delta \phi)}}
\end{equation}

In general $r_1>>h$ and therefore the Kerr parameter standard deviation scales as $\braket{\Delta a} \gtrapprox \frac{r_1^3}{2 \omega L r_s h\sqrt{M N}}$.%, it is better than measuring the ``two-way" time delay in the Michelson interferometer which scales as $\braket{\Delta a} \gtrapprox \frac{r_1^4}{\sqrt{M N}}$. 

%Also for the Schwarzschild radius we have the QCR bound:

%\begin{equation}
%\frac{\braket{\Delta r_s}}{r_s}\ge \frac{1}{ \omega r_s (\frac{L h}{r_1 r_2}+\frac{a h (r_1+r_2)}{r_1^2 r_2^2} )\sqrt{M H(\phi)}}
%\end{equation}

A larger height difference $h$ or length $L$ reduces the noise limit. For coherent probe states undergoing linear phase evolution, $H(\phi)=|\alpha|^2=N$. Therefore, we have the standard quantum noise limit $\propto \frac{1}{\sqrt{N}}$ as expected for coherent probe states. By using non-classical squeezed states the noise scales as $\frac{1}{N}$, known as the conventional Heisenberg limit \cite{BOI, zwierz} or with $\chi$ Kerr non-linearities the noise can scale as $\frac{1}{N^{3/2}}$ \cite{BOI07, KISH}.

\subsection{Mach-Zehnder interferometer}
\label{mz}
%\tikzstyle{block} = [draw, fill=blue!20, rectangle, 
%minimum height=3em, minimum width=6em]
%\tikzstyle{sum} = [draw, fill=blue!20, circle, node distance=1cm]
%\tikzstyle{input} = [coordinate]
%\tikzstyle{output} = [coordinate]
%\tikzstyle{pinstyle} = [pin edge={to-,thin,black}]
\begin{figure}
	\begin{center}
\includegraphics[scale=1]{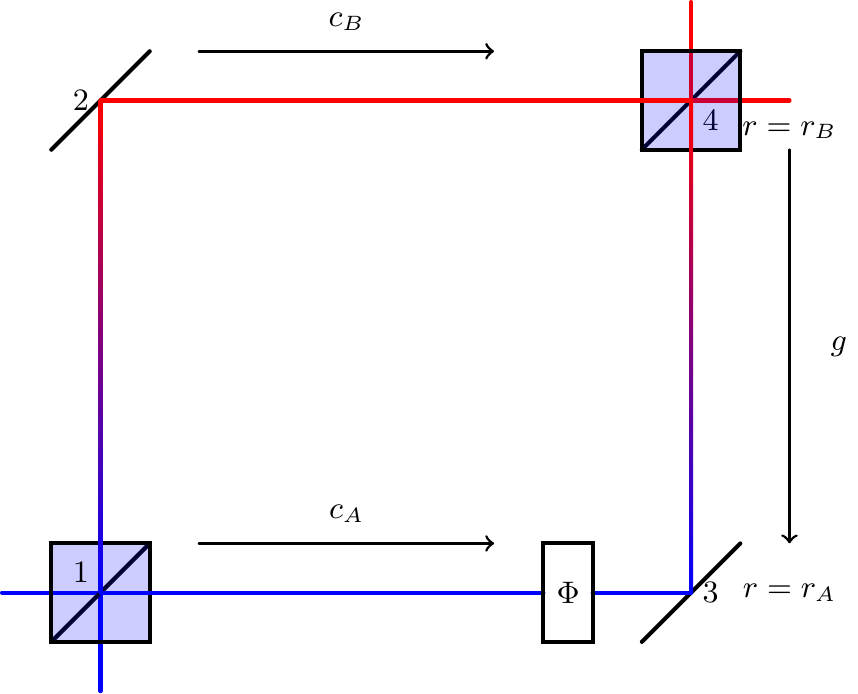}
	\end{center}
	\caption{A Mach-Zehnder interferometer of length $L$ and height $h$ stationary above the rotating planet. $\Phi$ is a phase shifter in the bottom arm to calibrate the interferometer to a dark port of zero intensity.}
	\label{int}
\end{figure}

	%The advantage of doing this is that we can directly measure the time delay $\Delta t_{Total} \approx 2 L v^2 = \frac{2 L r_s^2 a^2}{r^4}$ from the Kerr metric without the need to know $r_s$. 
	
	%We can measure the difference in phase between two counter-propagating beams of light using the Michelson interferometer as done in Section \ref{two}. However, this phase scales as $\propto \frac{a^2}{r^4}$ and the magnitude of the effect is too small to be detectable. 
	Let's consider a physical system that can detect the discrepancy in the velocity of light from the differential height effect in the Kerr metric. We consider a Mach-Zehnder interferometer (see Fig. \ref{int}) that is stationary with respect to the centre of mass of a rotating planet. We will work in far-away time coordinates. Although the final implications will be the same, this is an approach where no assumption is made about how the speed of light is measured locally. 
	
	The measured phase of the bottom arm of the Mach-Zehnder interferometer is $\Delta \phi_A=\omega \Delta t_A$ where $\omega$ is the frequency of light measured locally at the source and $\Delta t_A$ is the time as seen by a faraway observer, and $\Phi$ is a local phase shifter. At $r=r_A$ the faraway time $\Delta t_A=\frac{L}{c_A}$ where $c_A$ is the speed of light as measured by a faraway observer (see Eq. \ref{dxdt}) and $L$ is the arm length also seen by a faraway observer. We have set both arm lengths to be the same. Thus, in the top arm the phase is $\Delta \phi_B=\omega \Delta t_B=\omega \frac{L}{c_B}$. 

	\begin{figure}
\centering
\includegraphics[scale=0.4]{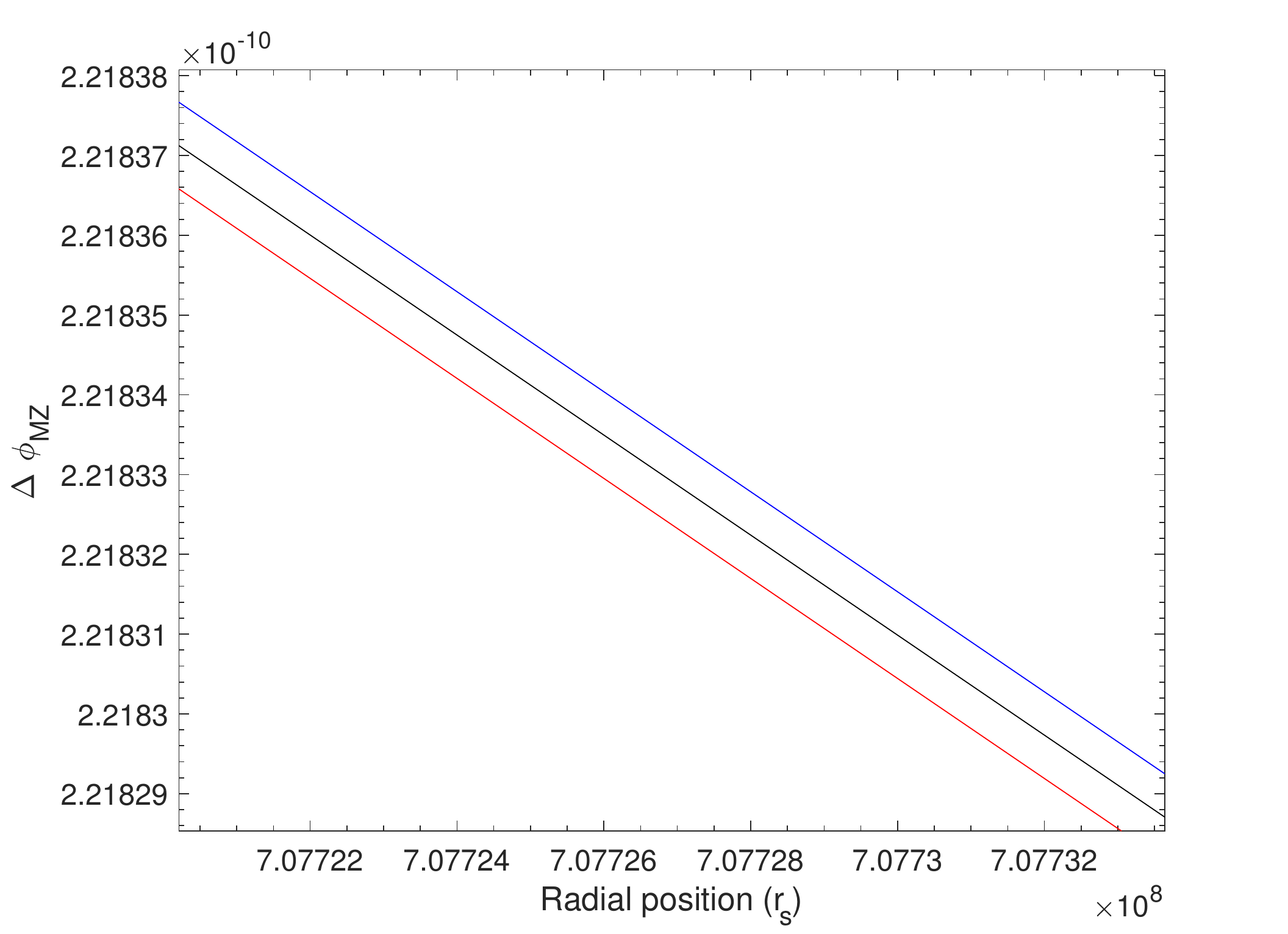}
\caption{Measured phase differences of $L=1$ $m$ and $h=1$ $m$ Mach-Zehnder interferometer in co- and counter- moving directions (blue and red respectively). Black line is in the Schwarzschild metric with $a=0$. We use the values for the Earth's Schwarzschild radius $r_s=9$ $mm$, rotation parameter $a=3.9$ $m$ and the operating frequency of light $\omega=k=2\times 10^{6}$ $m^{-1}$ corresponding to $500$ $nm$ measured locally at the source. }
\label{earth}
\end{figure}

We assume that $dr=0$ and the Mach-Zehnder interferometer arms are sufficiently small that the curvature is negligible. The tangential velocity of light depends on $R$ and the sign of $a$. The solution in the weak field limit is $c'=\frac{dx}{d t}=R \frac{d \phi}{dt} \approx 1\pm \frac{r_s a}{r^2}-\frac{r_s}{2 r}$. 	Where we have chosen the co-moving direction such that $c_{A} \approx 1 - \frac{r_s a}{r_{A}^2}-\frac{r_s}{2 r_A}$ and $c_{B} \approx 1 - \frac{r_s a}{r_{B}^2}-\frac{r_s}{2 r_B}$. The phase is thus:
	
	%The length of the arm should be the Lorentz contracted length in these coordinates where $\gamma=\frac{1}{\sqrt{1-u^2}}$ and $u$ is the relative velocity of the two observers assuming Geostationary Earth Orbit (GEO) i.e. $u=\Omega_E(R_B^2-R_A^2)\approx\frac{3 a}{5 R_A^2} (R_B^2-R_A^2)$. 

\begin{equation}
\begin{split}
\Delta &\phi_{MZ}-\Phi=\omega (\frac{L}{c_B}-\frac{L}{c_A})\\
&\approx \omega L ( (1 + \frac{r_s a}{r_{B}^2}+\frac{r_s}{2 r_B}+\frac{r_s^2 a}{r_B^3}) \\
&-(1 + \frac{r_s a}{r_{A}^2}+\frac{r_s}{2 r_A}+\frac{r_s^2 a}{ r^3_{A}}))\\
&\approx \omega L ( -\frac{r_s a h (2 r_A +h)}{r_A^4 (1+\frac{h}{r_A})^2} -\frac{h r_s}{2 r_A^2 (1+\frac{h}{r_A})} +r_s (\Omega_B-\Omega_A))
\end{split} 
\label{mzphase}
\end{equation}

	Where we have used the Taylor expansion $\frac{1}{1-x-y} \approx 1+x+y+2 x y$. Note that $\Omega_{A,B}=\frac{r_s a}{r_{A,B}^3}$. We have made the approximations $\frac{r_s}{r}<<1$, $\frac{a}{r}<<1$ and $h<<r_A$. Note that for the vertical arms, the accumulated phases are equal $\Delta \phi_{12}=\Delta \phi_{34}$ implying that there is no contribution to the total output phase. 

	We note that on Earth scale the effect of the Kerr rotation parameter is small. If we use the values for the Earth's Schwarzschild radius $r_s=9$ $mm$, rotation parameter $a=3.9$ $m$ and radius $r_B=6.37 \times 10^6$ $m$, and take the area of the interferometer as $A=L \times h=1$ $m^2$ and the operating frequency of light $k=2\times 10^{6}$ $m^{-1}$ (wavelength of $500$ $nm$) then the order of magnitude of the dominant term for the Kerr rotating effect is 
\begin{equation}
|\Delta \phi_{Kerr}|\approx \frac{2 k r_s a L h}{r_B^3} \approx 5 \times 10^{-16} 
\end{equation}	
	
	Conversely, the Schwarzschild time dilation effect is of the order $\Delta \phi_{Schwarzschild}=\frac{\omega L h r_s}{2 r_A r_B}=2.2 \times 10^{-10}$. Note that the term $\omega L r_s (\Omega_B-\Omega_A)\approx \omega L \frac{3 r_s a h}{r_B^4}\approx 10^{-22}$ is too small and can be neglected in further calculations. %Similarly, the term $\omega L(\frac{a^2 r_s^2}{2 r_B^4}-\frac{a^2 r_s^2}{2 r_A^4})\approx \frac{4 k L a^2 r_s^2 h}{r_B r_A^4} \approx 1 \times 10^{-30}$ can be neglected in further calculations. 
	
		%We give a simple argument for the angle dependence of the Mach-Zehnder phase as it is rotated along its centre. We have a velocity component parallel to the equator $a \cos{\beta}$. We have determined in Eq. \ref{speedp} (See Appendix \ref{proper}) that perpendicular to the equator, we have only the contribution from the Schwarzschild time dilation, and none from the Kerr metric (to an approximation). Thus, the Kerr phase simply becomes $\Delta \phi_{Kerr}\approx \frac{k r_s a \cos{\beta} L h}{r_B^3}$. 
	
	\emph{MZ interferometer calibration}. We set the total phase shift $\Delta \phi_{MZ}=0$ and thus the phase shifter $\Phi$ balances the interferometer to the dark port. Isolating the Kerr phase around the dark port is an optimal strategy for maximizing signal to noise ratio. We can see in Fig. \ref{earth} the phase of the interferometer if it were positioned in the co- and counter-moving directions. Thus, we can rotate the Mach-Zehnder interferometer with angle $\pi$ around its vertical axis and measure the $a$ sign dependence directly. Since only the sign of $a$ changes and $\Phi$ stays the same then we have,
	\begin{equation}
	\begin{split}
	\Delta' \phi_{MZ}-\Phi &\approx 2 \omega L ( \Omega_A r_A-\Omega_B r_B-r_s (\Omega_B-\Omega_A)) \\
	& \approx 2 |\Delta \phi_{Kerr}|
	\end{split}
	\end{equation}
	
	 Therefore, we have a signal which only depends on $a$. Without the anisotropy of light speed, there would be no signal and the phase would remain a dark port.

%What's the significance of $a=2 r_s$? Why does the effect dissapear?
%Clearly, as $a^2>> 2 r_s a$ then we have the dominant phase shift $\frac{L a^2 (r_B^2-r_A^2)}{2 r_A^2 r_B^2}$. Also, the relativistic effects could be entirely cancelled out if $(a^2-2 r_s a)\frac{(r_B^2-r_A^2)}{2 r_A r_B} +r_s h=a^2-2 r_s a+\frac{2 r_s h r_A r_B}{r_B^2-r_A^2}=0$

%Thus $a=\frac{2 r_s \pm \sqrt{4 r_s^2-8\frac{r_s h r_A r_B}{r_B^2-r_A^2}}}{2}= r_s \pm \sqrt{ r_s^2-2\frac{r_s r_A r_B}{r_B+r_A}}$

\section{Zero Angular Momentum Observer metric}
\label{zamo}
The co- and counter-propagating null light geodesics differ in the Kerr metric. However, locally we expect observers to isotropically measure $c=1$. It would be useful to transform to a reference frame in which the cross terms $d\phi dt$ vanish and where locally we obtain a flat space-time metric with $c=1$ \cite{taylor}. To determine this transformation, we consider the Killing vectors $\partial_t$ and $\partial_\phi$ that are responsible for two conserved quantities along the geodesic. These are the energy:
\begin{equation}
E= -k_\mu u^\mu=- g_{t \mu} u^\mu=-p_t=(1-\frac{r_s}{r}) \frac{d t}{d \tau}+\frac{r_s a }{r} \frac{d \phi} {d \tau}
\end{equation} 
And the angular momentum:
\begin{equation}
\mathcal{L}=g_{\phi \mu} u^\mu=-\frac{r_s a}{r} \frac{d t}{d \tau}+r^2\frac{d \phi}{d \tau}
\end{equation}
When we set $\mathcal{L}=0$ then we have that $\frac{d\phi}{dt}=\frac{r_s a}{r^3}$. Thus there remains an angular motion even with zero angular momentum. The interpretation here is that the rotating space-time drags an object close to the rotating mass, as seen by a far-away observer. 
If we are co-rotating in the zero angular momentum reference frame $d\phi'= d\phi_{ring}+\Omega dt$ with angular velocity $\Omega=\frac{r_s a}{r^3}$ then the metric cross terms $d\phi dt$ cancel out and becomes:

\begin{equation}
\begin{split}
ds^2&=-(1-\frac{r_s}{r})dt^2+r^2 d\phi^2_{ring}
\end{split}
\end{equation}
%\\&=(1-\frac{r_s}{r}+\frac{r_s^2 a^2}{r^2 R^2})dt^2-dx_{ring}^2
This is known as the zero angular momentum observer (ZAMO) metric \cite{taylor}. Taking $dt_{ring}=\sqrt{1-\frac{r_{shell}}{r}} dt$ we have that 
\begin{equation}
ds^2=dt_{ring}^2-r^2 d\phi_{ring}^2
\end{equation}

giving a locally flat metric for the ringriders in which $c=1$. 

We seek the metric in stationary shell coordinates: 
\begin{equation}
ds'^2=dt^2_{s}-r_{shell}^2 d\phi_{ring}^2
\end{equation}
where obviously again $c=1$ locally. 

However, there is a lack of simultaneity between events in the shell metric and events in the ring-rider metric (and hence faraway events). This is the source of the anisotropy of the speed of light. We have from the Lorentz transformation that a space-like event implies $dt_{ring}=\gamma (dt_s-v dx_s)=0$ where $v=\Omega r$ and $dx_s=r_{shell} d\phi_{ring}$, thus $dt_s=v r_{shell} d\phi_{ring}$. 

From the equivalence of the line elements we have 
\begin{equation}
\begin{split}
ds^2&=ds'^2\\
-r^2 d\phi_{ring}^2&=v^2 r_{shell}^2 d\phi_{ring}^2-r_{shell}^2 d\phi_{ring}^2
\end{split}
\end{equation} 
Therefore the ring-rider radius and stationary observer radius are equivalent $r=r_{shell}$.  

We have redefined the coordinate times of the respective ring-riders as the Schwarzschild time $d\tau=\sqrt{1-\frac{r_s}{r}} dt$. Between ring-riders, we have the usual Schwarzschild time dilation, as expected. The advantage of the ring-rider frame is that we can use Lorentz transformations to the stationary observer frame to determine the much more significant height differential effect.

\begin{figure}
\includegraphics[scale=1]{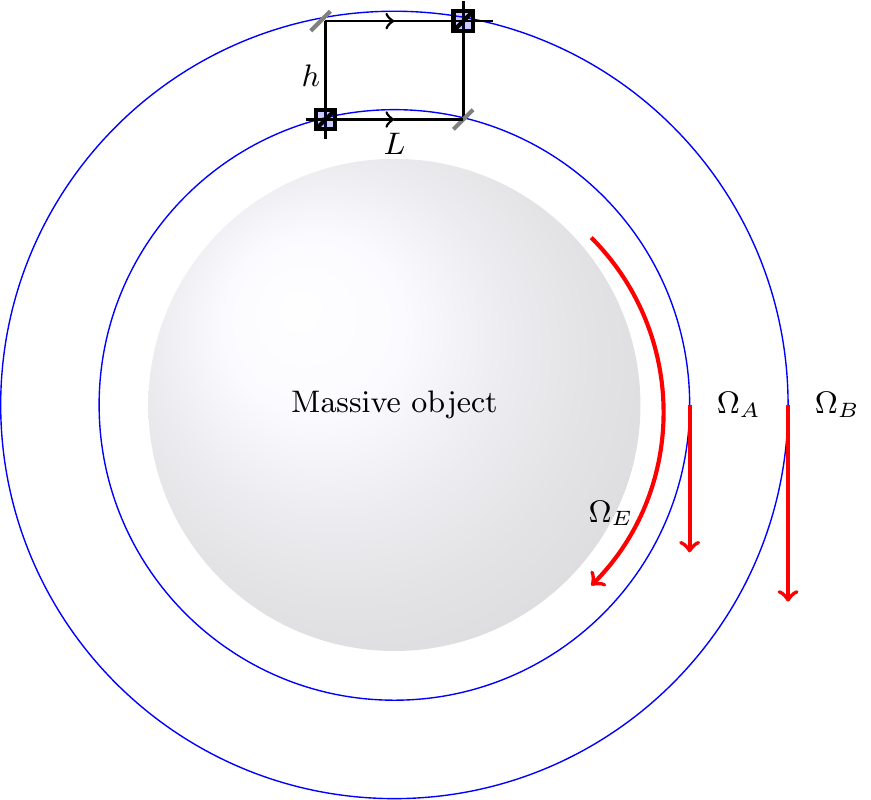}
\centering
\caption{A Mach-Zehnder interferometer of length $L$ and height $h$ stationary above the rotating massive object in the Kerr metric. Zero angular momentum ring-riders (blue) will have a locally flat space-time with $c=1$. Their angular frequency as seen from a far-away observer (red) are given by $\Omega_A=\frac{r_s a}{r_A^3}$ and $\Omega_B=\frac{r_s a}{r_B^3}$.}
\label{ringrider}
\end{figure}

\section{Ring-rider perspective}
\label{ringriderper}
We have previously shown that in the ZAMO flat metric the speed of light is $c=1$. It is helpful in understanding the physics of our estimation protocols to consider them from the perspective of ring-rider observers. This is also a convenient method to generalize to non-stationary interferometers. 
\subsection{Stationary Mach-Zehnder above rotating massive object}
%\label{station}
%Let's consider the Mach-Zehnder interferometer being lowered from far-away $R=\infty$ to the radial distances $R_A$ and $R_B$ above a rotating massive object. In this way, it will remain stationary. 
%We have already calculated the time delay using far-away coordinates between the travel times of light. This is given by:

%\begin{equation}
%\Delta t_{MZ}= L ( \Omega_B R_B-\Omega_A R_A-\frac{r_s h}{2 r_A r_B}+r_s (\Omega_B-\Omega_A))
%\label{farawaytime}
%\end{equation}

%Where we have assumed the anisotropic nature of the light in the Kerr metric. 

%\subsection{Far-away coordinates}
%{\color{red} Refer to MZ section and motivate reason we are considering this situation. Why ring-rider frame?}

%We assume that the Mach-Zehnder interferometer remains stationary for the entire travel time of the light. Thus only the effect of the Kerr metric affects the light. 
%The velocity of light as seen by a far-away observer is $c'\approx (1-\frac{r_s}{2r}\pm \frac{r_s a}{r R}+\frac{v_B^2}{2})$. Using the Schwarzschild time coordinates, the arm lengths as measured by light travelling at $c'_A$ and $c'_B$ are $ \frac{L}{\sqrt{1-\frac{r_s}{r_A}}}=c'_A t_A$ and $L=c'_B t_B \sqrt{1-\frac{r_s}{r_B}}$. Thus we want the faraway time difference 

Let's consider the Mach-Zehnder interferometer from the reference frames of the ring-riders. The ring-riders are in the flat metric (see Fig. \ref{ringrider}). Therefore, for each ring-rider, we can use the Lorentz Transformations. We maintain for now the weak field approximations that $\frac{a}{R}<<1$ and $\frac{r_s}{r}<<1$ such that the Mach-Zehnder interferometer is far enough away from the centre of the massive body.  Taking into account special relativity, a stationary observer would measure the travel time of light:
\begin{equation}
\begin{split}
t'_1&=\gamma (t_1+v_A x_A)=\gamma (L+v_A L)\\
&=\sqrt{\frac{1+v_A}{1-v_A}} L \approx (1+v_A) L
\end{split}
\end{equation}

Where $v_A=\Omega_A r_A$ is the relative velocity between the ring-rider and stationary observer at $r_A$ and $t_1=L$ is the travel time in the ZAMO flat metric. Note that the stationary observer as seen by the ring-rider is travelling in the negative $x$ direction. Similarly, for the ring-rider at $R_B$, $t'_2=\sqrt{\frac{1+v_B}{1-v_B}} L \approx (1+v_B) L$ where $v_B=\Omega_B r_B$.
For an observer at $r=\infty$, we use the coordinate times of the ZAMO metric. Since the coordinate times are 
\begin{equation}
t''_1= \frac{t_1'}{\sqrt{1-\frac{r_s}{r_A}}}\approx (1+\frac{r_s}{2 r_A}) L (1+v_A)
\end{equation}
and 
\begin{equation}
t''_2=\frac{ t_2'}{ \sqrt{1-\frac{r_s}{r_B}}}\approx (1+\frac{r_s}{2 r_B}) L (1+v_B)
\end{equation}
Thus the time delay is
\begin{equation}
\begin{split}
\Delta t&=t''_2-t''_1=L ((1+\frac{r_s}{2 r_B}) (1+\Omega_B r_B)\\
&-(1+\frac{r_s}{2 r_A})(1+\Omega_A r_A)) \\
&\approx L (\Omega_B r_B-\Omega_A r_A-\frac{r_s h}{2 r_A r_B}+\frac{r_s}{2} (\Omega_B -\Omega_A))
\end{split}
\end{equation}
These calculations are equivalent with using the null geodesics obtained from using the Kerr Metric in far-away coordinates in Eq. \ref{mzphase}. 
%{\color{red}
\subsection{Michelson interferometer}
Given that the far-away observer sees an anisotropic speed of light it is instructive to ask why a local Michelson interferometer fails to see an effect. A stationary observer sends a light beam tangential to the equator that bounces off a mirror $L$ distance away and returns to the observer. The time delay in this signal arm would be:
\begin{equation}
\begin{split}
\Delta t_{Signal}&=\frac{L}{\sqrt{1-\frac{r_s}{r}}} (1+v)\\
&+\frac{L}{\sqrt{1-\frac{r_s}{r}}} (1-v)\\
&\approx 2 L (1+\frac{r_s}{2r})
\end{split}
\end{equation}
The reference arm perpendicular to the equator is approximately the Schwarzschild local time as found in Eq. \ref{speedp} (see Appendix \ref{proper}). This is the same phase as the signal arm $\Delta t_{Ref} \approx 2 L (1+\frac{r_s}{2r})$. Thus the total phase difference is $\Delta \phi_{Michelson}=0$, implying that the speed of light is $c=1$ locally and isotropic, as expected from the special theory of relativity. From the point of view of the far-away observer, although the speed of light is anisotropic, they find the ``two-way" speed, to the mirror and back, is the same in each direction, leading to no phase shift. It may seem a contradiction with the results of the height differential effect, which requires $c$ to be anisotropic to see a signal in the MZ interferometer. However, this is due to a difference in the amount of space-time dragging at different radial positions in the Kerr metric that the MZ interferometer measures non-locally. %The ZAMO ring-rider and stationary observer both conclude that the ``two-way" velocity of light in their region of space is $c=1$. 

\subsection{Non-stationary co-moving observers on Earth}
\label{moving1}
%In the ZAMO metric, we previously used the transformation $d\phi_{ring}+\Omega dt$. However, for a non-stationary observer co-moving with the Earth, we have $d\phi_{ring}-\Omega_O dt$ where $\Omega_O$ is the observers angular velocity. Thus the Kerr metric becomes:
%
%\begin{equation}
%\begin{split}
%ds^2&=-(1-\frac{r_s}{r}-v\frac{2 r_s a}{r R}-\Omega^2 R^2)dt^2\\
%&-2R(v+\Omega R) d\phi dt+R^2 d\phi^2_{ring}
%\end{split}
%\end{equation}
%Thus solving the geodesic, we have
%   \begin{equation}
%   \begin{split}
%    \frac{d x}{dt}&=\frac{r_s a}{r \sqrt{r^2+a^2(1+r_s/r)}}+v\pm \sqrt{1-\frac{r_s}{r}}
%    \end{split}
%   \end{equation}
%
%Where $v_A=\Omega_O R$. 

In an experiment conducted say on Earth, the rotation of the non-stationary Earth observers must be taken into account. Our previous calculations have considered only a stationary Mach-Zehnder interferometer with the Earth rotating beneath. However, let's consider the bottom arm of the MZ interferometer on Earth's surface with the tangential velocity $v'_A=\Omega_E r_A-\Omega_A r_A$ and the top arm co-moving at  $v'_B=\Omega_E r_B-\Omega_B r_B$ with the same angular velocity $\Omega_E$ of Earth. This relative velocity between observers introduces an additional time dilation.

Using the Lorentz transformations, a stationary observer observer would measure the travel time of light at $r_A$:
\begin{equation}
\begin{split}
t'_1&=\gamma (t_1+v_A x_A)=\gamma (L+v_A' L)=\sqrt{\frac{1+v_A'}{1-v_A'}} L \\
&\approx (1+v_A'+\frac{v_A'^2}{2}) L
\end{split}
\end{equation}

Similarly, for the moving observer at $r_B$: 
\begin{equation}
t'_2=\sqrt{\frac{1+v_B'}{1-v_B'}} L \approx (1+v_B'+\frac{v_B'^2}{2}) L
\end{equation}
 
For an observer at $r=\infty$, we use the coordinate times of the ZAMO metric, $t''_A= \frac{t_1'}{\sqrt{1-\frac{r_s}{r_A}}}$ and $t''_B= \frac{t_2'}{\sqrt{1-\frac{r_s}{r_B}}}$. Thus:
%\begin{equation}
%\begin{split}
%\Delta t&=L(\frac{1}{\frac{r_s a}{r_B R_B}+v_B+\sqrt{1-\frac{r_s}{r_B}}}-\frac{1}{\frac{r_s a}{r_A R_A}+v_A+\sqrt{1-\frac{r_s}{r_A}}}) \\
%&\approx (1-\frac{r_s a}{r_B R_B}+\frac{r_s}{2 r_B}-v_B-\frac{v_B^2}{2}-\frac{r_s v_B}{2 r_B} - 1\\
%&+\frac{r_s a}{r_A R_A}-\frac{r_s}{2 r_A}+\frac{r_s v_A}{2 r_A} +v_A+\frac{v_A^2}{2})\\
%&= \Delta t_{MZ}-\Omega_E h L-\frac{\Omega_E^2 h L (R_A+R_B)}{2}
%\end{split}
%\label{moving}
%\end{equation}
\begin{equation}
\begin{split}
\Delta t&=t''_B-t''_A\\
&=L ((1+\frac{r_s}{2 r_B}) (1+v_B'+\frac{v_B'^2}{2})\\
&-(1+\frac{r_s}{2 r_A})(1+v_A'+\frac{v_A'^2}{2})) \\
&\approx L (\frac{r_s h}{2 r_A r_B}+v_B'-v_A'+\frac{r_s v_B'}{2 r_B}-\frac{r_s v_A'}{2 r_A} ) \\
& \approx \Delta t_{MZ}+\Omega_E h L+\frac{\Omega_E^2 h L (2 r_A +h)}{2}
\end{split}
\label{moving}
\end{equation}
%  %Where $\Delta \phi=\Delta \phi_{Stationary}+$

Where we have neglected the terms $(\Omega_A r_A)^2$ and $(\Omega_B r_B)^2$. The term $\Omega_E h L$ is a classical effect due to the relative motion of the observers but the term $\frac{\Omega_E^2 h (2 r_A+h) L}{2}$ is the higher order correction due to special relativity. We calibrate the MZ interferometer such that the total phase $\Delta \phi_{MZ}=0$ and then we rotate it. The only remaining terms in Eq. \ref{moving} are linear with the rotation. Thus the new phase is 

\begin{equation}
\Delta \phi'_{MZ}=2 \Delta \phi_{Kerr}+2\omega_0 \Omega_E h L
\end{equation}

The Kerr phase varies inversely with $r^3$, and thus in principle can be distinguished from the classical effect. However, let's consider uneven lengths of the interferometer such that the classical term cancels. Thus we have $r_A L_A=r_B L_B$, and the Kerr phase is 
\begin{equation}
\begin{split}
|\Delta\phi_{Kerr}|&\approx\frac{\omega_0 L_B r_s a}{r_B^2}-\frac{\omega_0 L_A r_s a}{r_A^2}\\
&= \omega_0 r_s a (\frac{L_A r_A}{r_A^3(1+\frac{h}{r_A})^3}-\frac{L_A}{r_A^2}) \\
&\approx \frac{3\omega_0 L_A h r_s a}{r_A^3}
\end{split}
\end{equation}
We note that the vertical phases $\Delta \phi_{12}$ and $\Delta \phi_{34}$ are not equal to each other. However, since we rotate the Mach-Zender interferometer through $\pi$ then $\Delta' \phi_{12}=\Delta \phi_{34}$ and $\Delta' \phi_{34}=\Delta \phi_{12}$. Thus the phase difference (given calibration to the dark port before rotation) at the output has no contribution from the phases of the vertical arms.

%We also have a coupling with the Schwarzschild metric $\frac{\Omega_E r_s L}{2}$.

%\section{Numerical estimate of phase}
%
%{\color{red} is this a repeat?}
%Now we want to determine the feasibility of measuring $a$ using a compact size interferometer.
%To measure the Kerr parameter $a$, we can set up the Mach-Zehnder interferometer and obtain the phase shift $\Delta \phi=\omega_0 \Delta t_{MZ}$. We can rotate the Mach-Zehnder interferometer and measure the $a$ sign dependence directly. 
%
%Using a reference frequency as measured by a faraway observer, the total phase will be $|\Delta \phi|= \omega_0 L ( \Omega_B R_B-\Omega_A R_A-\frac{r_s h}{2 r_A r_B}+\frac{r_s}{2} (\Omega_B-\Omega_A))\approx\omega_0 L(\frac{r_s a h}{r_B^3}+\frac{r_s h}{2 r_A r_B}+\frac{3 r_s a h}{2 r_B^4})$. The largest contribution from the Kerr rotation will come from the term $\Delta \phi_{Kerr}=\omega_0 L \frac{r_s a h}{r_B^3} \approx 3 \times 10^{-16} $ for an a $1$ $m$ $\times$ $1$ $m$ interferometer with the central frequency $\omega_0 = k_0= 2 \times 10^6$ $m^{-1}$. In comparison with the Schwarzschild term, $\Delta \phi_{Schwarzschild}=5 \times 10^{-10}$. Note that the term $\omega_0 L \frac{3 r_s a h}{2 r_B^4}\approx 10^{-22}$ is  too small and can be neglected.
\begin{figure}
\centering
\includegraphics[scale=0.4]{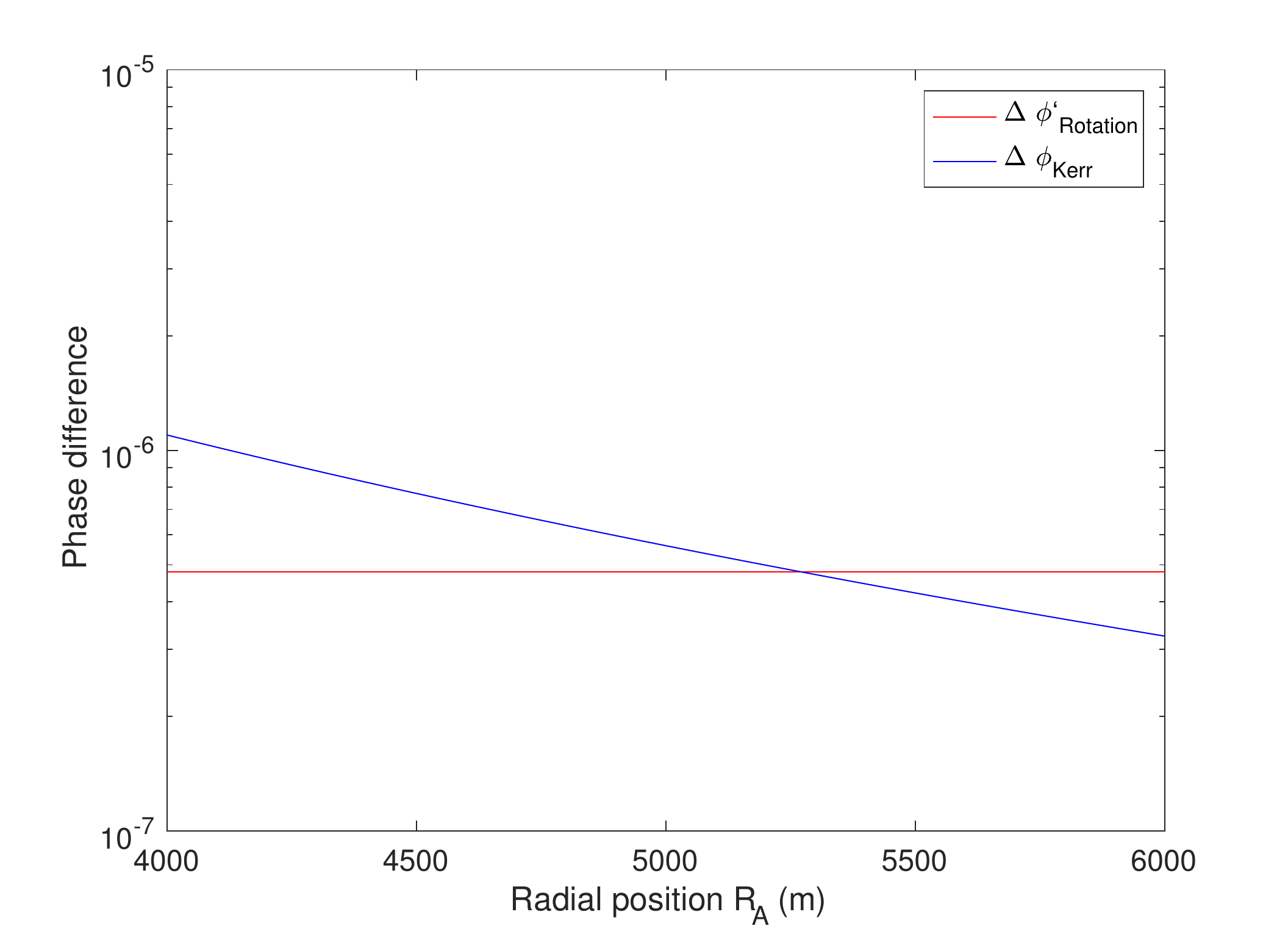}
\caption{Measured phase differences of $L=1$ $m$ and $h=1$ $m$ Mach-Zehnder interferometer around the radial position at which the Kerr phase (blue) becomes dominant on Earth compared to the phase due to the classical rotation (red).}
\label{graph1}
\end{figure} 

\begin{figure}
\centering
\includegraphics[scale=0.4]{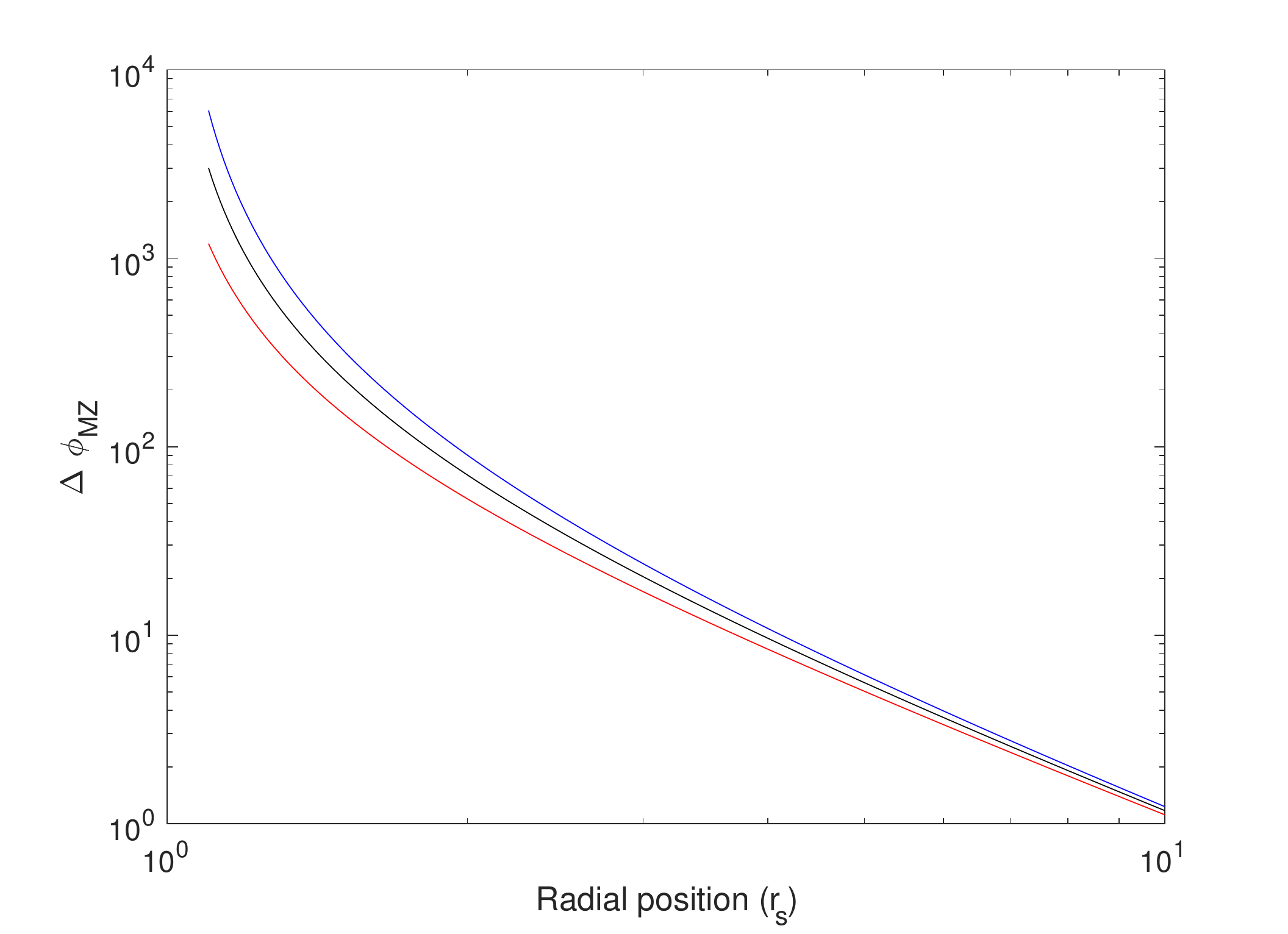}
\caption{Measured phase differences of $L=1$ $m$ and $h=1$ $m$ Mach-Zehnder interferometer near a black hole of Schwarzschild radius $r_s=10$ $km$, angular momentum $a=\frac{r_s}{8}$ and the operating frequency of light $k=2\times 10^{6}$ $m^{-1}$. Here we have the MZ phases for co-moving (red), counter-moving (blue) and no rotation $a=0$ (black).}
\label{blackhole}
\end{figure}

\subsection{Probing the Kerr phase on Earth using MZ interferometer}

An interesting calculation is to estimate how compact an object with Earth mass and spin would need to be such that the Kerr term was dominant over the effect of the spin. The relative velocity term is $|\Delta \phi_{Rotation}|=\omega_0 L\Omega_E h\approx 5 \times 10^{-7}$ for a fixed interferometer with the angular frequency of the Earth $\Omega_E=\frac{7.2 \times 10^{-5}}{c}$ $Hz$. To determine $a$, we need to isolate it from the dominant effect of Earth's rotation. 

We can vary the position of the interferometer while keeping its size constant. The contribution from the rotation term $\Delta \phi_{Rotation}\approx \omega_0 \frac{\Omega_E h L}{2}$ is approximately constant. We want to determine at what radial position the Kerr effect becomes dominant. This occurs when $\Delta \phi_{Kerr} \geqslant \Delta \phi_{Rotation}$. Therefore, $\omega_0 L \frac{r_s a h}{r_B^3}=\omega_0 L \Omega_E h$ which implies that $r_B=(\frac{r_s a}{\Omega_E})^{1/3}\approx 5$ $km$. Note that the condition $\frac{a}{r_A}<<1$ is still satisfied. In Fig. \ref{graph1},  we have the same interferometer over a range of positions extending $2$ $km$ around the point at which the Kerr phase becomes significant. Clearly an Earth bound measurement is very far from this condition. However, for a compact object such as a neutron star of the same Schwarzschild radius it is possible in principle. 

%The Kerr effect would be $\Delta \phi_{Kerr}=1 \times 10^{-8}$ and the Schwarzschild effect is $\Delta \phi_{Scharzschild}=7.2 \times 10^{-3}$.

%\section{Quantum Resources}
%The parameter is encoded in the phase $\Delta \phi(a, r_s)=\omega_0 L ( \Omega_B R_B-\Omega_A R_A-\frac{r_s h}{2 r_A r_B}+\frac{r_s}{2} (\Omega_B-\Omega_A))\approx \omega_0 L(\frac{r_s a h}{r_B^3}-\frac{r_s h}{2 r_A r_B})$. The lower bound is given by:
%\begin{equation}
%\frac{\braket{\Delta a}}{a} \ge \frac{1}{a \sqrt{M H(\phi) (\frac{d \phi}{d a})^2}}
%\end{equation}
%
%For a linear phase estimation using coherent probe states, it is known that $H(\phi)=|\alpha|^2=N$. Thus:
%
%\begin{equation}
%\frac{\braket{\Delta a}}{a} \ge \frac{R_B^3}{\omega_0 L a r_s h \sqrt{M N}}
%\end{equation}
%
%Therefore $\frac{\braket{\Delta a}}{a} \ge \frac{10^{16}}{\sqrt{M N}}$ and since normal repetition rates of $M=1$ $GHz$ are possible then $N \geqslant 10^{22}$ with 1\% precision for a $1$ $m$ $\times$ $1$ $m$ interferometer considering a perfect channel with no loss. %The standard deviations of $a$ using squeezed coherent probe states or non-linear Kerr medium are $\frac{\braket{\Delta a}}{a} \ge \frac{10^{16}}{\sqrt{M N^2}}$ and $\frac{\braket{\Delta a}}{a} \ge \frac{10^{16}}{\sqrt{M N^3}}$ respectively  \cite{BOI07, KISH}. This implies beyond SQL scaling. 

\section{Extremal Black Holes}
\label{extremal}

To explore the strong field situation, let's now lower our stationary Mach-Zehnder interferometer close to a black hole. We can no longer use the approximations $\frac{a}{r}<<1$ and $\frac{r_s}{r}<<1$. We must use the full solution of $c_A$ and $c_B$ of the unapproximated Kerr metric as in Eq. \ref{kerrmetric} and calculated in Appendix \ref{extremebh}. We note that the Kerr metric is a good description for a collapsed black hole, but not for the exterior metric of neutron stars \cite{pappas}.
We can see in Fig. \ref{blackhole} for a black hole of Schwarzschild radius $r_s=10$ $km$ and angular momentum $a=\frac{r_s}{8}$, the phase difference for a co- (red) and counter- (blue) direction Mach Zehnder interferometer. The two directions of the Mach-Zehnder interferometer become increasingly distinguishable as it gets closer to the event horizon at $r=r_s$. 

\section{Conclusion}
We have determined the quantum limits of estimating the Kerr parameter which arises from the anisotropy of the speed of light. %An effect arises for light returning from a mirror in the metric. We can see this ``two-way" effect in a Michelson interferometer and the phase is proportional to $\frac{a^2}{r^4}$. 
We propose a stationary Mach-Zehnder interferometer that can directly measure the Kerr parameter $a$ direction dependence. We identify the flat metric where the ring-rider velocity of light is locally $c=1$. We find the same Kerr phase using Lorentz transformations between stationary and ring-riders in this ZAMO flat metric. Also, we find that the ``two-way'' velocity of light is isotropic and $c=1$ as measured by a Michelson interferometer. However, our Mach-Zehnder interferometer is no longer a dark port after it is rotated by $\pi$ because of the combined effect of the anisotropy of light and the difference in the amount of space-time dragging in the radial position. On Earth, we have to consider non-stationary observers which adds an additional classical phase that dominates the Kerr phase. Using a variation on the Mach-Zehnder set-up can cancel this additional classical phase with only the Kerr phase remaining.

	Recent experiments using microwave resonators have been able to detect the anisotropy of light with a precision of $\Delta c/c\approx 10^{-17}$ \cite{herm}. Our Mach-Zehnder interferometer predicts a change in the speed of light due to the Kerr metric of $\Delta c_{Kerr}/c=\frac{h a r_s}{r^3}\approx 10^{-20}$. In principle, future devices need only increase precision by $3$ orders of magnitude to measure the Kerr phase on a small scale Mach-Zehnder interferometer. Using coherent probe states, the noise of the phase is the standard noise limit (SNL) $\Delta \phi \ge \frac{1}{\sqrt{M N}}$. For $M=10$ $GHz$ measurements \cite{ghz}, this suggests that $N=10^{22} - 10^{26}$ per light pulse. This would imply extremely high power, which is one of the current limiting factor to increasing phase sensitivity. 

\section*{Acknowledgements} 
We thank Jorma Louko for pointing out an error in the original manuscript and also Robert Mann and Daiqin Su for useful discussions. This work was supported in part by the Australian Research Council Centre of Excellence for Quantum Computation and Communication Technology (Project No. CE110001027) and financial support by an Australian Government Research Training Program Scholarship.
%{\color{blue} Ultimately, using the Mach-Zehnder interferometer to probe the Kerr phase could be within reach of future tabletop experiments.}

\appendix
\section{Proper length perpendicular to the equator}
\label{proper}
Let's consider the proper length perpendicular to the equator. The Kerr metric away from the equator is \cite{visser}:
\begin{equation}
ds^2=-(1-\frac{r_s r}{\Sigma}) dt^2+\frac{\Sigma}{\Delta} dr^2+\Sigma d\theta^2
\end{equation}

Where $\theta$ is the azimuth in spherical coordinates, and $\Sigma=r^2+a^2 \cos{\theta}^2$.

Therefore, we set $dt=0$ and $dr=0$ and get the proper distance $d\sigma=\sqrt{r^2+a^2 \cos^2{\theta}} d\theta$. However, for a massive planet, in the weak field limit, we have $d\sigma =r \sqrt{1+\frac{a^2}{r^2} \cos^2{\theta}} d\theta \approx r d\theta$. The velocity of light is given by solving the null geodesic for the weak field Kerr metric:
\begin{equation}
\begin{split}
ds^2=0&=-(1-\frac{r_s r}{r^2+a^2 \cos^2{\theta}}) dt^2+(r^2+a^2 \cos^2{\theta}) d\theta^2 \\
&\approx -(1-\frac{r_s}{r}) dt^2+d\sigma^2
\end{split}
\end{equation}
And thus the time travelled by light is:
\begin{equation}
\Delta t_{Normal}=\frac{L}{\frac{d\sigma}{dt}}=\frac{L}{\sqrt{1-\frac{r_s}{r}}}\approx 2 L (1+\frac{r_s}{2 r})
\label{speedp}
\end{equation}
Which is the same as in the Schwarzschild metric.
\section{Extremal black holes}
\label{extremebh}
Let's consider the full solution to the speed of light without any weak field approximations. The phase is therefore:

\begin{equation}
\Delta \phi=\omega (t_B-t_A)=k L (\frac{1}{c_B}-\frac{1}{c_A})
\end{equation}

\begin{figure}
\centering
\includegraphics[scale=0.6]{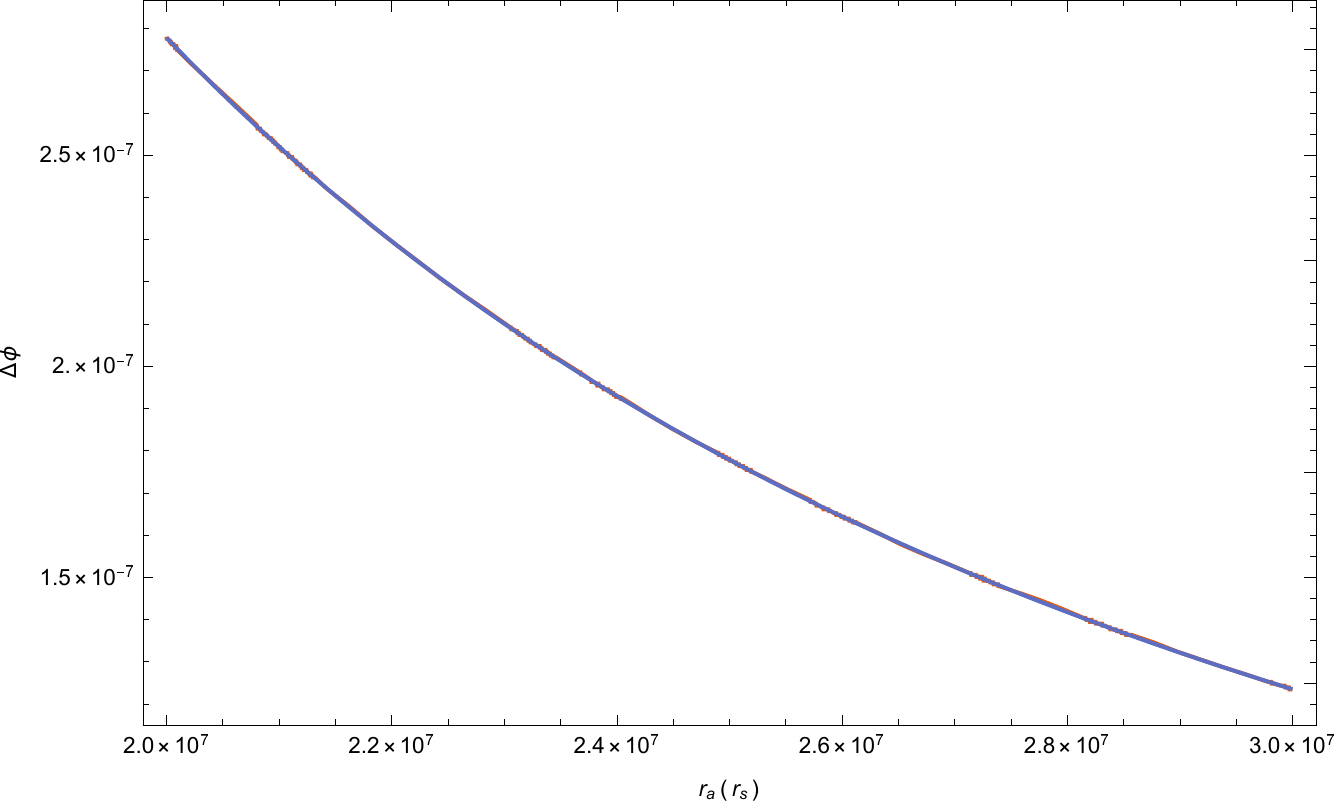}
\caption{Difference in exact phase as determined numerically for full solution of $c$ (red) and weak field approximation (blue) for Earth parameters. (Note that $r_s=9$ $mm$, $h'=111$ and $a'=433$)}
\label{graph2}
\end{figure}

Where $c_B=\frac{r_s a}{r_B \sqrt{r_B^2+a^2(1+r_s/r_B)}} \pm \sqrt{\frac{r_s^2 a^2}{r_B^2(r_B^2+a^2(1+r_s/r_B))}+(1-\frac{r_s}{r_B})}$. Using units of $r_s$, $a\rightarrow a' r_s$, $r_A \rightarrow r'_A r_s$ and $r_B \rightarrow r'_B r_s$. This simplifies to $c_B=\frac{1}{r'_B \sqrt{ \frac{r_B'^2}{a'^2}+(1+1/r'_B)}} \pm \sqrt{\frac{1}{r_B'^2((\frac{r_B'^2}{a'^2}+(1+1/r'_B))}+(1-\frac{1}{r'_B})}$. Let's consider the values of an almost extremal black hole with $r_s=10$ $km$, $a'=\frac{1}{8}$ with $r'_B=r'_A+h'$ where $h'=\frac{1}{10000}$ since $h=1$ $m$. We can see in Fig. \ref{graph1} the phase difference for the full solution of $c$ (red) and the weak field approximation (blue) for this extremal black hole. The weak field approximation obviously fails near the event horizon. However, for Earth parameters $r_s=9$ $mm$, $h’=111$ and $a’=433$ representing $h=1$ $m$ and $a=3.9$ $m$, there is no difference between the exact solution for $c$ and the weak field approximation on the Earth’s surface (see Fig. \ref{graph2}).

\begin{figure}
\centering
\includegraphics[scale=0.7]{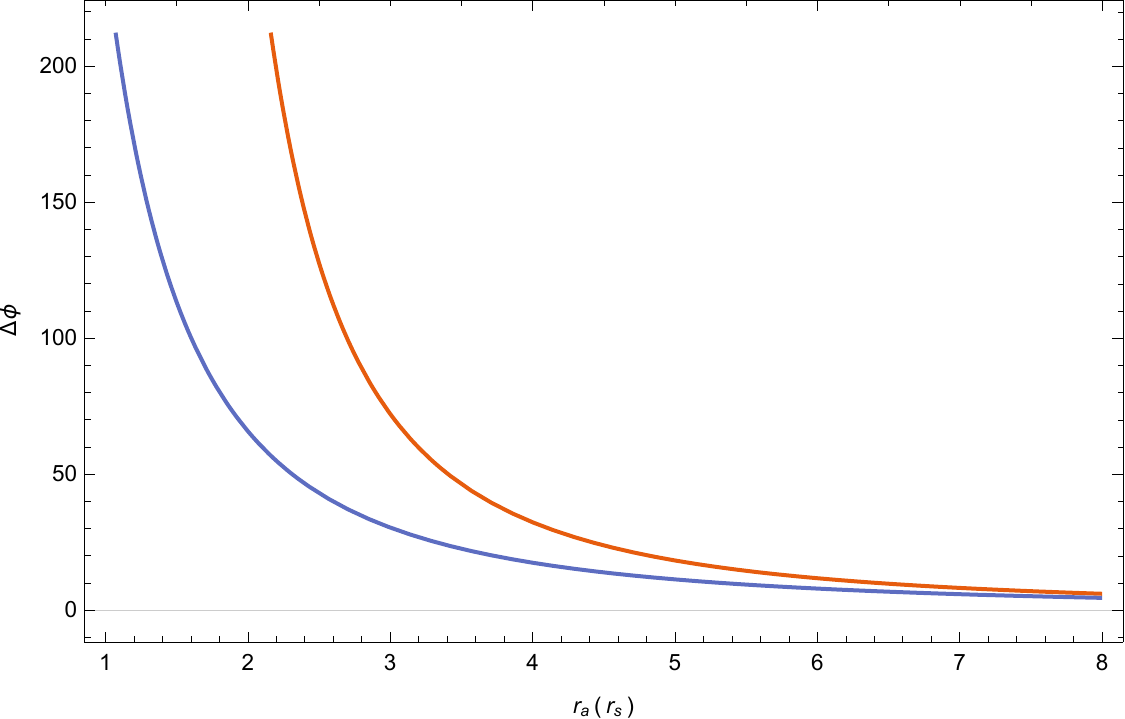}
\caption{Difference in exact phase as determined numerically for full solution of $c$ (red) and weak field approximation (blue). (Note the extremal black hole parameters $r_s=10$ $km$, $h'=10^{-4}$ and $a'=\frac{1}{8}$)}
\label{graph1}
\end{figure}

%\section{Exact solution to radial interferometer}
%
%Let's consider the phase for the radial interferometer:
%
%\begin{equation}
%\Delta \phi_{Radial}=\omega (t_B-t_A)=k S_A (\frac{R_B}{R_A c_B \sqrt{1-\frac{1}{r_B'}}}-\frac{1}{c_A \sqrt{1-\frac{1}{r_A'}}})
%\end{equation}
%
%Note that the dominant term is $\Delta \phi_{Radial} \propto S_A (\frac{R_B}{R_A}-1)$.
%
%\begin{figure}
%\centering
%\includegraphics[scale=0.6]{exactradial5.eps}
%\caption{Difference in phase of radial interferometer as determined numerically for full solution of $c$ (red), the weak field approximation (blue) and the non-rotating Schwarzschild case (orange). (Note the extremal black hole parameters $r_s=10$ $km$, $h'=10^{-4}$ and $a'=\frac{1}{8}$)}
%\label{graph}
%\end{figure}
%
%\begin{figure}
%\centering
%\includegraphics[scale=0.6]{minusdominant.eps}
%\caption{Difference in phase of radial interferometer excluding dominant term $\frac{h}{r_A}$ as determined numerically for full solution of $c$ (red), the weak field approximation (blue) and the non-rotating Schwarzschild case (orange). (Note the extremal black hole parameters $r_s=10$ $km$, $h'=10^{-4}$ and $a'=\frac{1}{8}$)}
%\label{graph}
%\end{figure}

\end{document}